\documentclass[12pt,preprint]{aastex}		

\shortauthors{Oliveira et al.}
\shorttitle{New Members of Serpens}

\begin{document}

\title{New X-ray Selected Pre-Main Sequence Members of the Serpens Molecular Cloud}

\author{ Isa Oliveira\altaffilmark{1,4}, Margriet van der
  Laan\altaffilmark{2}, \& Joanna M. Brown\altaffilmark{3} }
\altaffiltext{1}{McDonald Observatory, The University of Texas at
  Austin, 1 University Station, C1402, Austin, TX 78712, USA; email:
  oliveira@astro.as.utexas.edu} 
\altaffiltext{2}{Leiden Observatory, Leiden University, P.O. Box 9513,
  2300 RA Leiden, The Netherlands}
\altaffiltext{3}{Harvard-Smithsonian Center for Astrophysics, 60
  Garden St., MS 78, Cambridge, MA 02138, USA}
\altaffiltext{4}{Harlan J. Smith Postdoctoral Fellow}

\begin{abstract}

  The study of young stars no longer surrounded by disks can greatly
  add to our understanding of how protoplanetary disks evolve and
  planets form. We have used VLT/FLAMES optical spectroscopy to
  confirm the youth and membership of 19 new young diskless stars in
  the Serpens Molecular Cloud, identified at X-ray wavelengths.
  Spectral types, effective temperatures and stellar luminosities were
  determined using the optical spectra and optical/near-IR
  photometry. Stellar masses and ages were derived based on PMS
  evolutionary tracks. The results yield remarkable similarities for
  age and mass distribution between the diskless and disk-bearing
  stellar populations in Serpens. We discuss the important
  implications these similarities may have on the standard picture of
  disk evolution.

\end{abstract}


\section{Introduction}
\label{sintro}

The manner in which protoplanetary disks evolve and, potentially, form
planets is one of the most exciting subjects of research in
astrophysics in the last decade. Through multi-wavelength
observations, combined with a growing number of models and theories,
many complex processes have been found to play a role in explaining
how planets could possibly form from the small dust and gas initially
present in those disks. It is not yet clear exactly under which
circumstances the different processes are determinant. However, it is
clear that the study of protoplanetary disks alone is unlikely to lead
to a full picture of disk evolution.

A powerful way to learn more about disks is to study young stars that
no longer are surrounded by them. It is believed that all stars are
born harboring circumstellar disks, through conservation of angular
momentum during the star formation process. Observations have shown
that not all disks have the same lifetime, but that they vary between
approximately 1 and 10 Myr, with about 50\% of dusty disks
disappearing within 3 Myr (e.g. \citealt{HA01,HE08}).  Differences
between young stars of the same age with and without disks could yield
a greater understanding of the processes that affect disk evolution,
and eventually allow time for planets to form.

The difficulty in studying young stars without disks lies in
identifying them. Young stars are most commonly found on the basis of
their colors, which takes into account the presence of the dusty disk
emission, prominent in the infra-red (IR) region of the system
(star+disk) spectrum. If a young star is no longer surrounded by a
disk, its optical-IR colors are indistinguishable from
fore-/background stars observed in the same direction as the
star-forming region to which it belongs. With broad-band optical/IR
photometric studies, it can be difficult to identify candidate
diskless young stars.

The picture changes considerably when studying a region at X-ray
wavelengths. In this short-wavelength regime, young stars are very
active whether or not a disk is still present
(e.g. \citealt{PR05,FE07}). Indeed, X-ray emission has been often used
to identify young stellar objects in the literature
(e.g. \citealt{FK81,WK81,FM99,GE05,PA06,GU07,WI07,TE07}). Confirmation
of their young stellar nature, however, is necessary through other
means, the most powerful of which being optical spectroscopy
(e.g. \citealt{WA88,SC08}).

The Serpens Molecular Cloud is a young star-forming region ($\sim$2
Myr, \citealt{OL13}). VLBI measurements by \citet{DZ10} of EC95 in The
Serpens Core indicate it at a distance of 415 pc. Given the size of
the cloud, it is reasonable to assume a distance of 415 $\pm$ 15 pc
for its entirety. The young stellar population still surrounded by
disks in Serpens beyond its core has been uncovered in the infrared by
the {\it Spitzer Space Telescope} Legacy Program `c2d'
\citep{HA06,HB07,HA07}. Unlike the Serpens Core, which has been well
studied in a variety of wavelengths (e.g. \citealt{WI07,WI09,WI10}),
nothing was known about the diskless young stellar population of
Serpens beyond its core until it was observed with {\it XMM-Newton}
(Brown et al., in prep). Two {\it XMM-Newton} pointings cover about
$2/3$ of the Serpens area observed by the c2d, and dozens of objects
were identified by Brown et al. as young diskless star candidates, for
which confirmation is still necessary.

In this paper we report on an optical multi-object spectroscopic
survey designed to confirm the youth and membership of those
newly-discovered young diskless star candidates discovered in Serpens
with {\it XMM-Newton}. Section \ref{sdata} describes the VLT/FLAMES
observations and data reduction. The data are analysed in Section
\ref{sanalysis}, where spectral types, stellar temperatures and
luminosities are determined. Spectral energy distributions (SEDs) are
constructed with the addition of optical, near- and mid-IR photometry
from the literature. Then, the objects are placed in a HR diagram
overlaid with pre-main sequence evolutionary tracks, from which
individual ages and masses are derived. In Section \ref{sdis}, the
bona-fide diskless young stellar population of Serpens is discussed
and contextualized with respect to the disk-bearing population
\citep{OL09,OL10,OL13}. Finally, we present our conclusions in Section
\ref{scon}.

\section{Observations and Data Reduction}
\label{sdata}

The spectroscopic data were obtained using the Fibre Large Array
Multi-Element Spectrograph (FLAMES) on the 8.2m UT2 (Kueyen) telescope
of ESO, in Paranal, Chile, in service mode in June, 2009. FLAMES
consists of three components: a Fibre Positioner (OzPoz), a
medium-high resolution optical spectrograph GIRAFFE, and a link to the
instrument UVES. The data used in this work consist of two observing
blocks from FLAMES/GIRAFFE using the MEDUSA fibers. This setting
allows for up to 132 fibers of $1.2\arcsec$ aperture in the
$25\arcmin$ diameter field of view of the instrument. The wavelength
range chosen, 6437--7183 \AA{}, offers a spectral resolution of 0.8
\AA{} and covers temperature-sensitive features that allow spectral
classification.

The two observed fields (AO5 and AO6) were centered on the {\it
  XMM-Newton} observations. An input list of the young stellar
candidates coordinates was provided based on X-ray detections from
Brown et al. (in prep.). AO5, observed by {\it XMM-Newton} on
2007-04-15 in program 0402820101, was centered at 277.254625
+0.49727778, and AO6, observed 2008-04-16 in XMM program 0503240201,
was centered at 277.34775 +0.8228333. {\it XMM-Newton} has a slightly
larger aperture than FLAMES at 30' versus 25' so candidates on the
edges of the XMM fields could not be observed. All X-ray sources,
regardless of optical or infrared characteristics were considered for
follow-up. Preference was given to targets with optical counterparts
which are not part of the {\it Spitzer} disk sample in the same region
\citep{HA06,HB07,HA07} and had no previously known spectral
information. Fiber placement was optimized using the FPOSS Fibre
configuration program of FLAMES, which permitted observation of most
candidates. However, some candidates could not be simultaneously
observed and had to be dropped. Unused fibers were placed on `blank'
positions of the sky for sky subtraction. To maximize dynamic range,
fields were observed using multiple 2000 second exposures. Field AO5
had 5 such exposures and field AO6 had 4. A summary of the
observational details can be found in Table \ref{t_obs}.

\begin{figure}[!h]
\begin{center}
\centering
\includegraphics[width=0.9\textwidth]{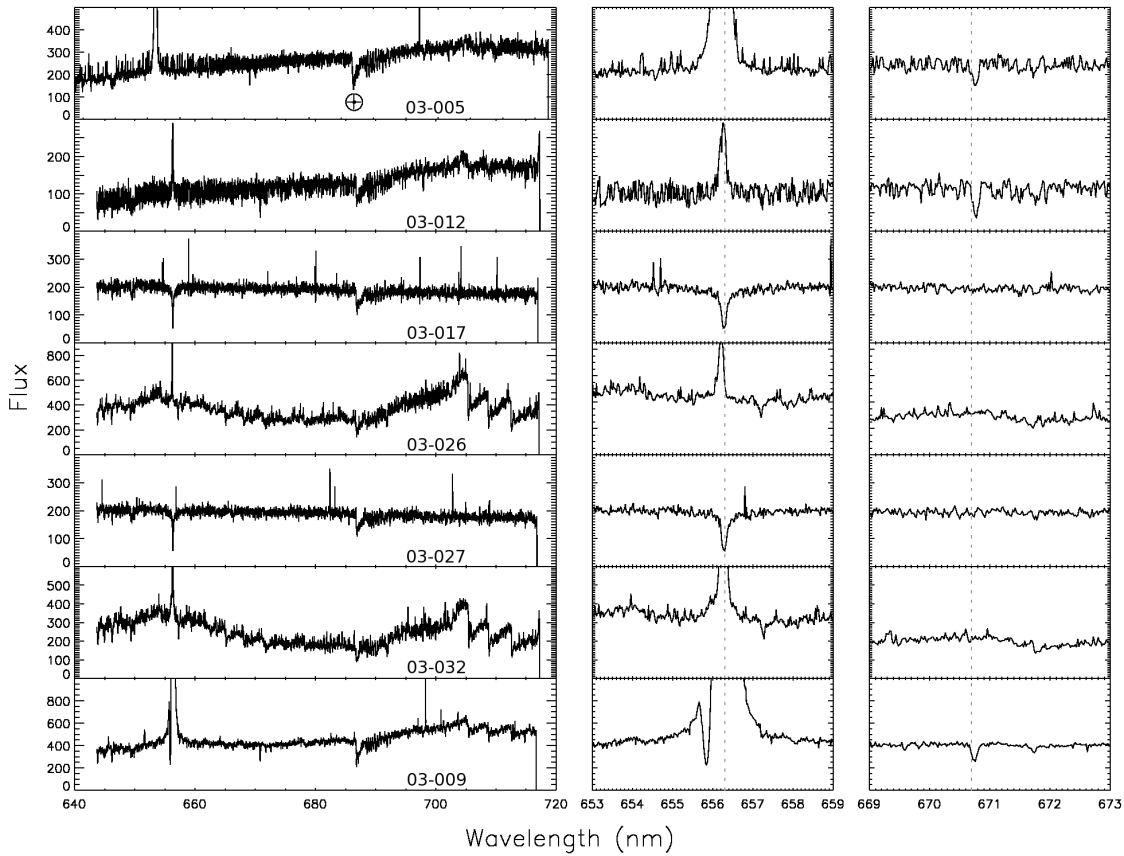}
\end{center}
\caption{Selected spectra of a representative sample of the observed
  objects in this sample. The left panel shows the entire wavelength
  coverage with its correspondent object ID. The most prominent
  features seen in late-type stars are the titanium oxide (TiO)
  absorption bands at 7050 -- 7150 \AA{}. The middle panel shows the
  H$\alpha$ region, and the right panel shows the Li {\sc i} line. }
\label{fspec}
\end{figure}

Data reduction was performed within the ESO reduction pipeline
GASGANO. The pipeline performs frame correction for detector defects,
fiber localization and tracing, flat-fielding and fiber transmission,
correction of scattered light and wavelength calibration. The science
spectra were extracted within IDL after wavelength calibration. For
each observed field, all sky spectra were combined with a $2\sigma$
clipping in order to exclude cosmic rays or any other artifacts. This
combined sky spectrum was then subtracted from the science spectra to
remove sky lines. Since each field was observed more than once (see
Table \ref{t_obs}), all observed spectra of the same object are
combined, also with a $2\sigma$ clipping, leading to the final
spectrum of each object. Figure \ref{fspec} shows a representative
sample of the spectra of the objects in this sample.

The final dataset consists of 86 observed objects, being 26 from field
AO5 and 60 from field AO6 (details in Tables \ref{t_radec5} and
\ref{t_radec6}, respectively). However, 61 of these objects turned out
to be much too faint and their spectra are underexposed, without
enough signal-to-noise ratio (S/N) to show any of the features
necessary for spectral classification (S/N $\gtrsim$ 10 per
pixel). These objects are marked as low S/N in Tables \ref{t_radec5}
and \ref{t_radec6} and are left out of further analysis.

\section{Data Analysis}
\label{sanalysis}

\begin{figure}[!h]
\begin{center}
\centering
\includegraphics[width=0.3\textwidth]{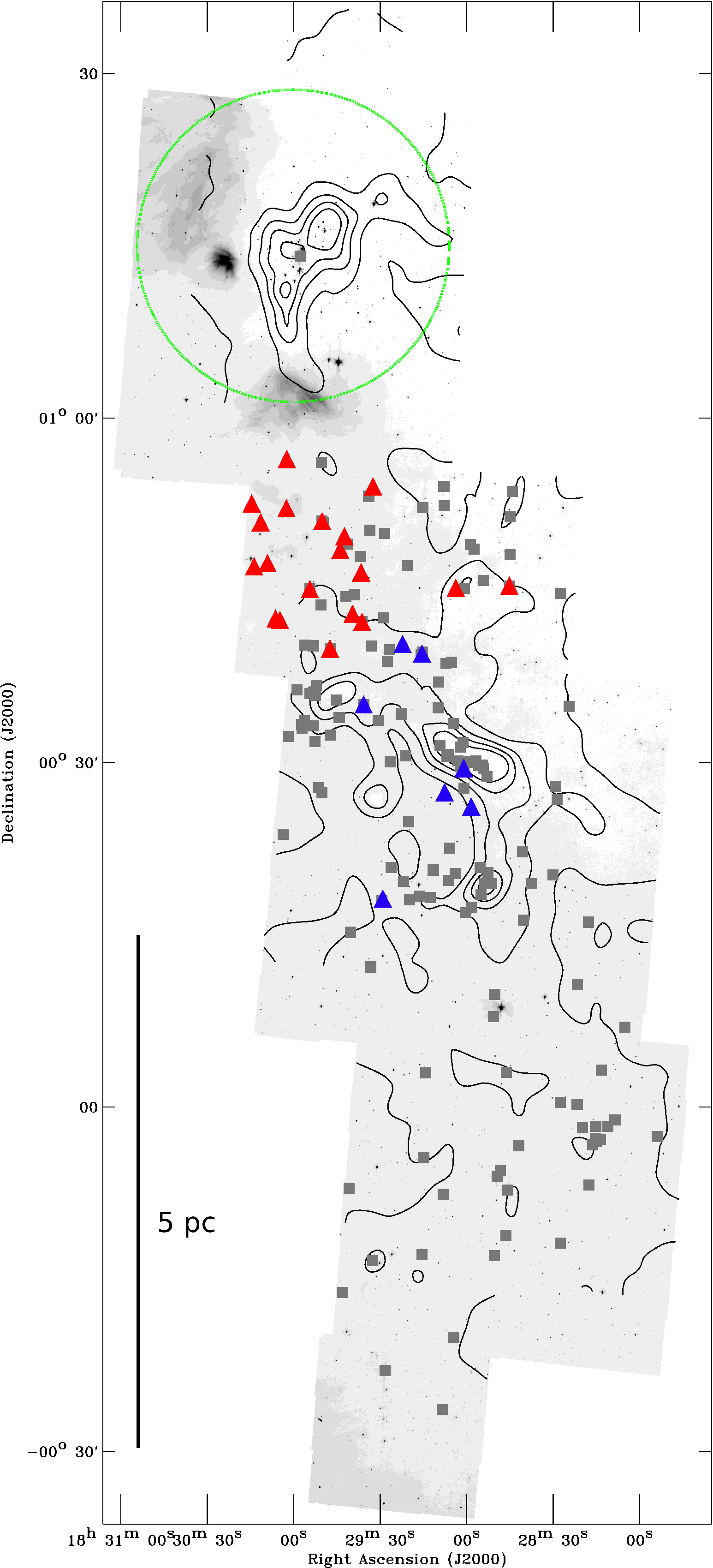}
\end{center}
\caption{Observed objects over-plotted on the 8.0 $\mu$m IRAC 4 map of
  the Serpens molecular cloud.  Young star candidates from this work
  are shown as red (field AO6) and blue (field AO5) triangles. Gray
  squares show the positions of the young stars with disks from
  \citet{OL10} for comparison. The contours (5, 10, 15, 20 and 25 mag)
  of visual extinction are derived from the c2d extinction maps
  \citep{EV07}. The large green circle indicates the Serpens Core.}
\label{serp}
\end{figure}

The positions of the 25 objects analyzed are shown in Figure
\ref{serp}. The 7 objects in field AO5 are shown in blue, while the 18
objects in AO6 are in red. The Serpens members with disks, as observed
with {\it Spitzer} by \citet{OL10} are shown in gray for comparison.

It is worth noting that the spectra presented here cover the H$\alpha$
line, an indicator of mass accretion onto the star
\citep{WB03,MU03,NA04}, and the Li {\sc i} absorption line, often used
as a general indicator of stellar youth. Lithium can be depleted in
low-mass stars (late-K and M-types) due to fusion in the stellar
interior. Evolutionary models predict depletion to start at $\sim10 --
20$ Myr and proceed quickly. Stars more massive than $\sim$ 1
M$_\odot$ develop radiative cores, limiting mixing to the center,
which retards depletion \citep{RA01,HA03}. Due to variations in S/N,
Li {\sc i} could not be detected in all of the spectra presented
here. The detection criterion adopted is an equivalent width greater
than 0.1 \AA{}. The presence (in emission or absorption) or absence of
these lines, as well as the presence of IR-excess in those sources,
are indicated in Table \ref{t_spt}.

\subsection{Spectral Types}
\label{sspt}

\begin{figure}[h]
\begin{center}
\includegraphics[width=0.4\textwidth]{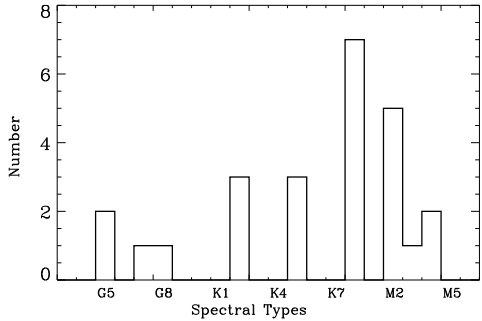}
\end{center}
\caption{Distribution of spectral types of young diskless star
  candidates in Serpens based on the classification scheme described
  in section \ref{sspt}. }
\label{f_spt}
\end{figure}

Spectral classification of the 25 objects is performed by comparing
each spectrum to a library of standards, following \citet{OL09} and
\citet{MO11}. The routine used here classifies the optical spectra by
a direct comparison with the grid of standards from EXPORT
\citep{MO01} and one from \citet{MO97}. In this routine the science
spectra are first normalized to the continuum and then over-plotted on
the normalized standards of different spectral types. A $\chi^2$
minimization is performed to find the best fit amongst all spectral
types, followed by a visual inspection of each object. This method
yields an accuracy of a half spectral class given the range of
standard spectra available. A larger uncertainty is found for the
spectra with lower S/N. The distribution of spectral types from this
work is shown in Figure \ref{f_spt}, while the results for individual
stars can be found in Table \ref{t_spt}.

Similar to the distribution of spectral types of the young stars with
disks \citep{OL09}, the majority of stars in this sample are of late K
and M spectral types, with only a few being earlier G-types.

\subsection{Stellar Temperatures}
\label{steff}

Effective temperatures ($T_{eff}$) for the individual objects were
determined using calibrations that relate them to spectral types. For
stars of type earlier than M0, the relationship established by
\citet{KH95} is used, while for stars of later type the relationship
from \citet{LU03} was adopted. The errors in $T_{eff}$ come directly
from the errors in the spectral types (Section \ref{sspt}). The
results can be found in Table \ref{t_res}.

\subsection{Stellar Luminosities}
\label{slum}

Stellar luminosities are calculated by integrating the NEXTGEN model
atmosphere \citep{HA99,AL00} corresponding to each object's spectral
type, scaled to its dereddened optical and near-IR photometry and
assuming a distance 415 $\pm$ 15 pc \citep{DZ10}. Optical R- and
Z-band photometry for this region is available in the literature
\citep{SP10}, although not all objects have been detected due to the
high extinction towards some regions of the cloud. When optical
photometry was not available, the Two Micron All Sky Survey (2MASS)
photometry was used to scale the model atmospheres. Similar methods
for luminosity estimates have been broadly used in the literature
\citep{KH95,VA97,AL08,ME08,OL09,OL13,MO11}. The uncertainties in
luminosity are derived from the errors in the distance and on the
extinction ($\pm 2$ mag). The results can be found in Table
\ref{t_res}.

Objects 03-009 and 03-027 have no available optical or near-IR
photometry and therefore are left out of any further analysis. All the
following discussion concerns the remaining 23 objects.

\subsection{Spectral Energy Distributions}
\label{ssed}

\begin{figure*}[!h]
\begin{center}
\centerline{\includegraphics[width=0.85\textwidth]{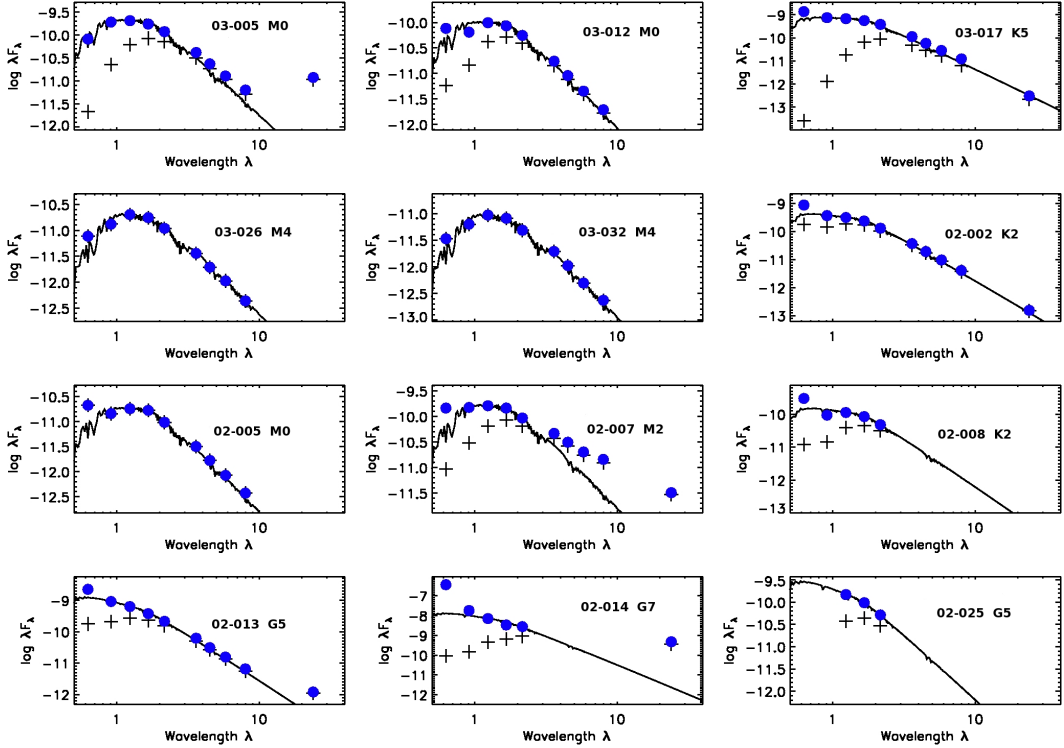}}
\end{center}
\caption{Spectral energy distributions of the new young diskless
  population in Serpens. Each SED has the corresponding object ID (as
  in Tables \ref{t_radec5} and \ref{t_radec6}). The solid black line
  indicates the NextGen Atmosphere model for the corresponding
  spectral type of each object, also shown in the plot. Plus signs
  indicate the observed photometry while solid blue circles denote the
  dereddened photometry. No SEDs are shown for the objects without
  photometric data (objects 03-009 and 03-027).}
\label{fsed1}
\end{figure*}

\begin{figure*}[!h]
\begin{center}
\centerline{\includegraphics[width=0.85\textwidth]{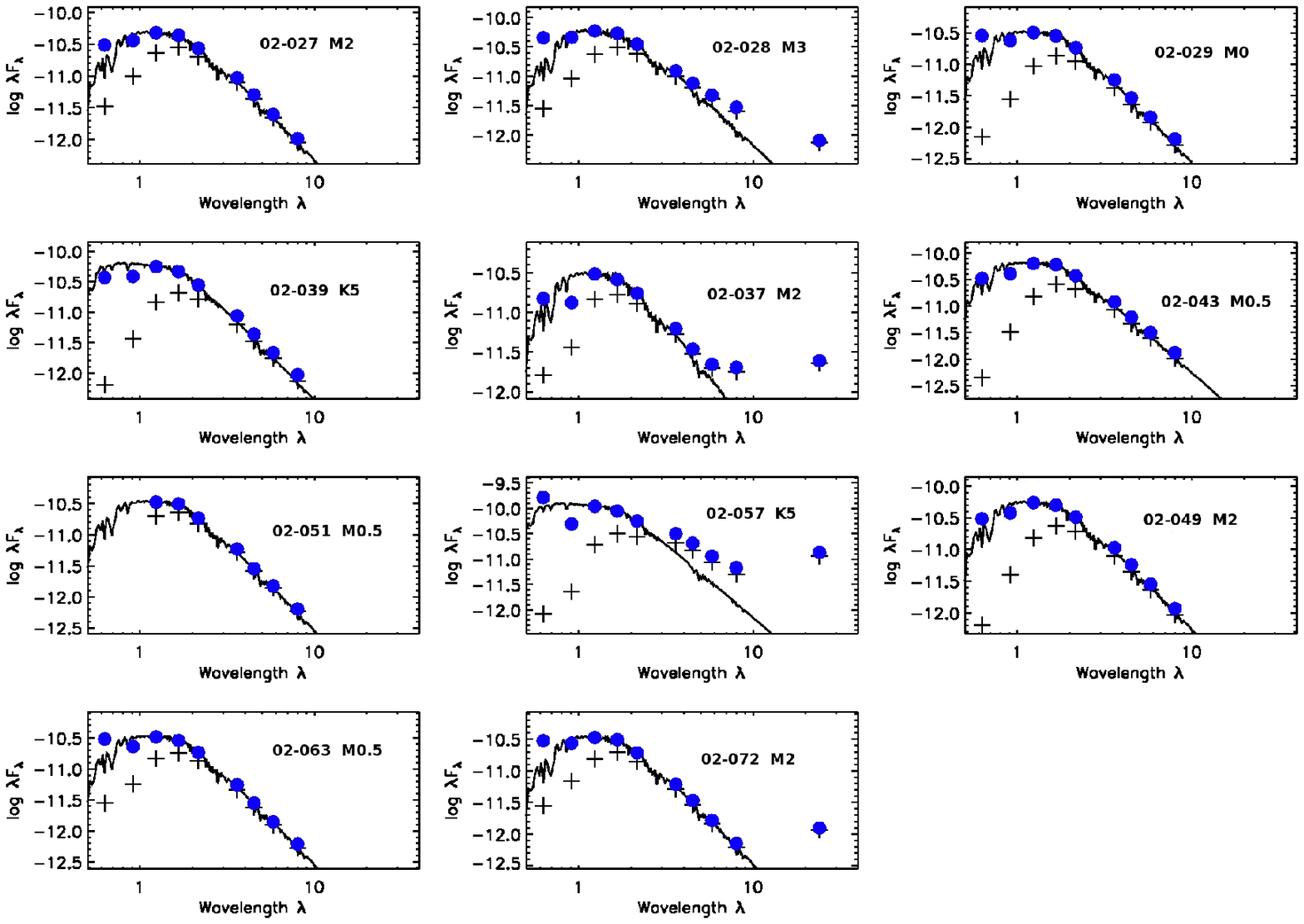}}
\end{center}
\caption{SEDs, continued.}
\label{fsed2}
\end{figure*}

The constructed spectral energy distributions (SEDs) for the 23 young
diskless star candidates are shown in Figure \ref{fsed1} and
\ref{fsed2}. Besides the corresponding NEXTGEN model atmosphere (solid
black line) and available optical/near-IR photometry from the
literature, as discussed in the previous section, {\it Spitzer} mid-IR
photometry is added where available. The IRAC (at 3.6, 4.5, 5.8 and
8.0 $\mu$m) and MIPS (at 24 $\mu$m) band fluxes were published by
\citet{HA06,HB07,HA07}. In the figures, the crosses indicate the
observed photometry, while the blue filled circles show the extinction
corrected photometry. Individual extinction values were derived to
best match the stellar photosphere, using the extinction law of
\citet{WD01}.

\subsection{Stellar Ages \& Masses}
\label{shrd}

\begin{figure*}[!h]
\begin{center}
\centerline{\includegraphics[width=0.5\textwidth]{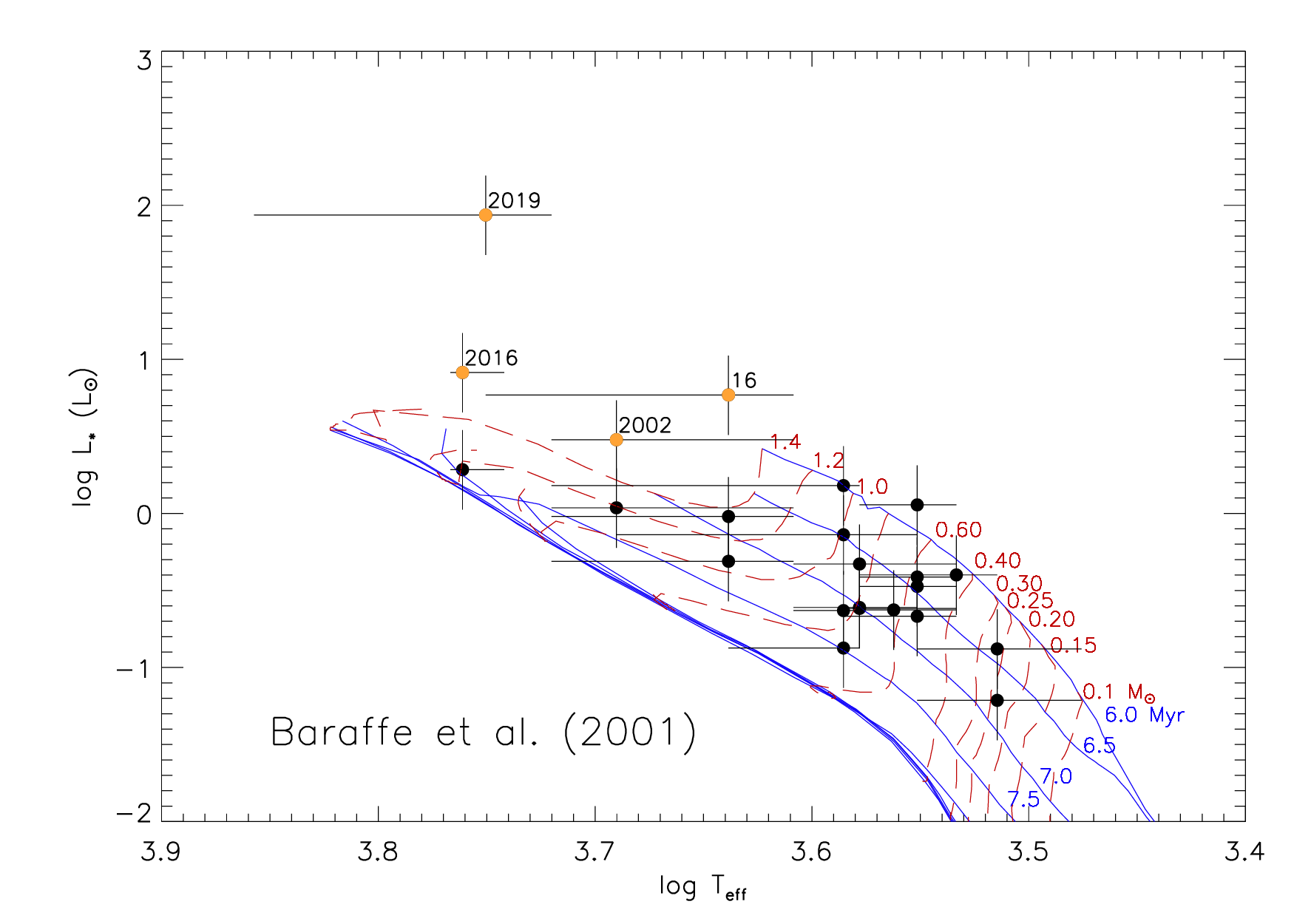}
\includegraphics[width=0.5\textwidth]{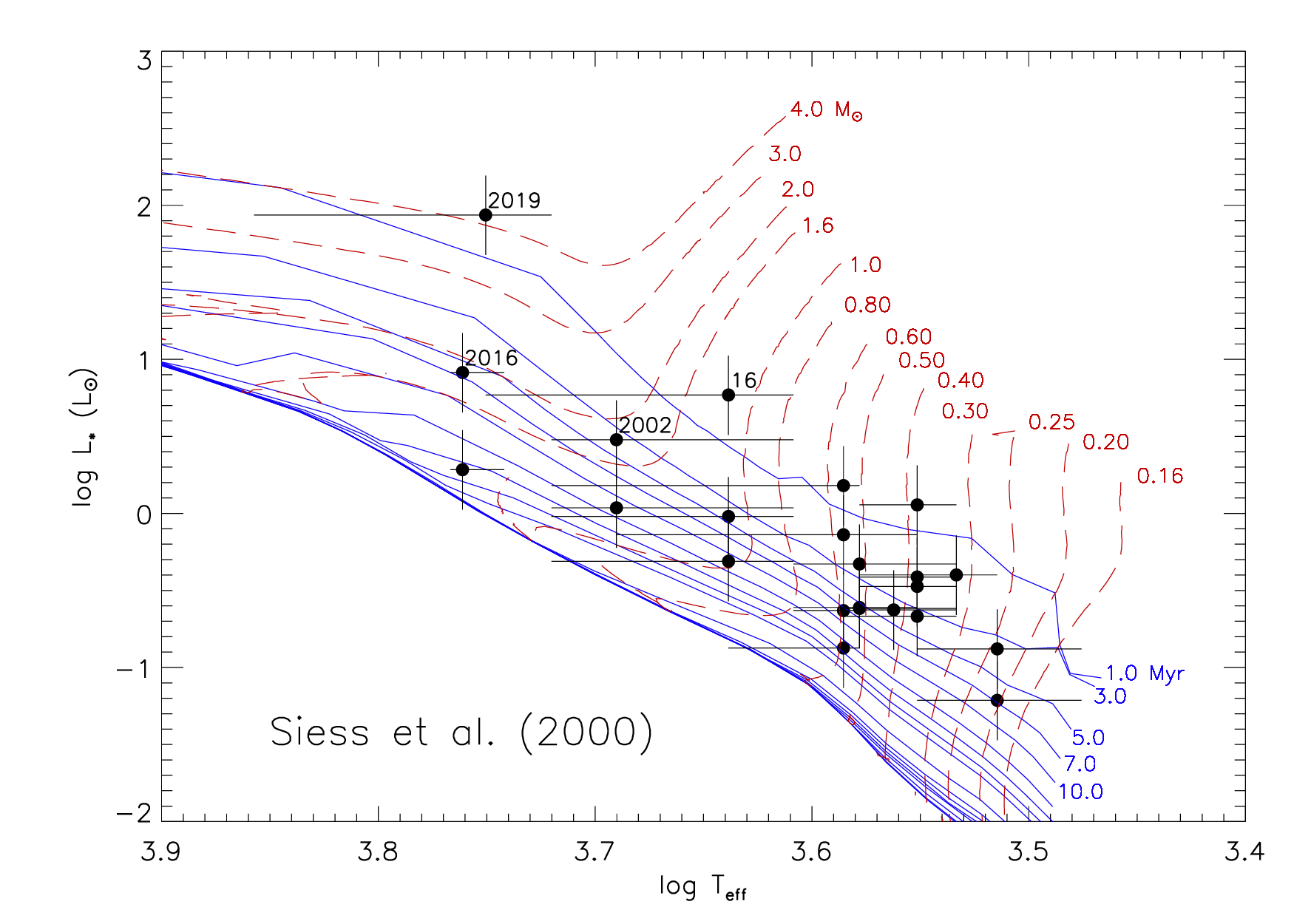}}
\end{center}
\caption{HR diagrams overlaid with the pre-main sequence evolutionary
  models of \citet{BA01}, left, and \citet{SI00}, right.}
\label{fhrd}
\end{figure*}

Once $T_{eff}$ and $L_{star}$ are known, it is possible to place the
objects in a HR diagram (Figure \ref{fhrd}). With the aid of pre-main
sequence (PMS) evolutionary tracks, ages and masses can be derived for
young stars by comparing each object's position to the isochrones and
mass tracks of a particular model. Due to the physics and calibration
of each model, the models of \citet{BA01} are used for stars less
massive than 1.4 M$_\odot$, while more massive stars are compared to
the models of \citet{SI00}. The results are displayed in Table
\ref{t_res}.

Due to the relatively old ages ($> 2 \times 10^7$ yr) derived for
objects 02-005, 02-008, 02-025 and 02-039, it is unclear whether they
indeed belong to the cloud, given the uncertainty of PMS evolutionary
models. It is worth noting that the spectra of objects 02-005 and
02-025 do not show Li {\sc i}, while the line is seen in the spectra
of 02-008 and 02-039. At this point it is difficult to confirm
membership and these four objects are left out of the further
discussion. The remaining 19 objects are confirmed new members of
Serpens.

\subsubsection{Infrared Excess}
\label{sire}

It can be seen from Figures \ref{fsed1} and \ref{fsed2} that 6 out of
19 new confirmed members of Serpens (03-005, 02-007, 02-028, 02-037,
02-057 and 02-072) show some IR-excess. Objects 02-007 and 02-028 show
excess emission throughout the {\it Spitzer} bands. This is considered
weak IR-excess for being lower than the median excess for objects in
Taurus \citep{FU06}, the prototype for full disks. The other four
objects have SEDs indicative of cold or transitional disks, where no
excess is seen in the near- to mid-IR, and a more substantial excess
is present at mid- to far-IR. Object 03-005 was indeed previously
confirmed as a cold disk with {\it Spitzer} IRS spectra \citep{BM10}.
Objects 02-037, 02-057 and 02-072 are new cold disk candidates. None
of the four older objects, not yet confirmed to be members of the
cloud, show any IR excess.

Following \citet{WB03}, we identify H$\alpha$ emission in Table
\ref{t_spt} as strong (full width at $10\%$ of peak intensity higher
than 270 km s$^{-1}$) or weak (H$\alpha$ 10\% < 270 km s$^{-1}$). In
addition, in Table \ref{t_spt} `A' indicates that H$\alpha$ is seen in
absorption. Three out of the 19 new members show strong emission of
the H$\alpha$ line. These objects (03-005, 02-037 and 02-072) also
show IR excess, indicating that disks are still present. None of the
confirmed diskless sources show strong H$\alpha$ emission, albeit some
do show weak H$\alpha$ emission, which can be attributed to
chromospheric activity \citep{WB03}.

This way, 86 stars were observed by FLAMES but only 25 objects are
bright enough for spectral classification. Out of these 25 objects
analyzed, 19 are confirmed new members of Serpens, while the other 6
could not be confirmed or rejected at this point. Out of the new
confirmed members of the cloud, 2/3 are show to be diskless,
while the other 1/3 show IR-excess weaker than the median of Taurus.\\

\section{Discussion}
\label{sdis}

The most straightforward explanation for why the disks around these
newly confirmed young stars have already dissipated would be that
these stars are systematically a few Myr older than the stars with
disks in the same area of Serpens \citep{OL13}. The higher ages would
be sufficient for their disks to have evolved and dissipated. This
could be the case for the older objects 02-005, 02-008, 02-025 and
02-039, if these objects indeed belong to the cloud. Another
possibility would be that these diskless stars are consistently more
massive than the disk-bearing stars in the same area of Serpens of the
same age. It has been shown that disk lifetimes are related to the
host star mass, with higher mass stars having disks that dissipated
faster \citep{CA06,KK09}.

\begin{figure*}[!h]
\begin{center}
\centering
\includegraphics[width=0.85\textwidth]{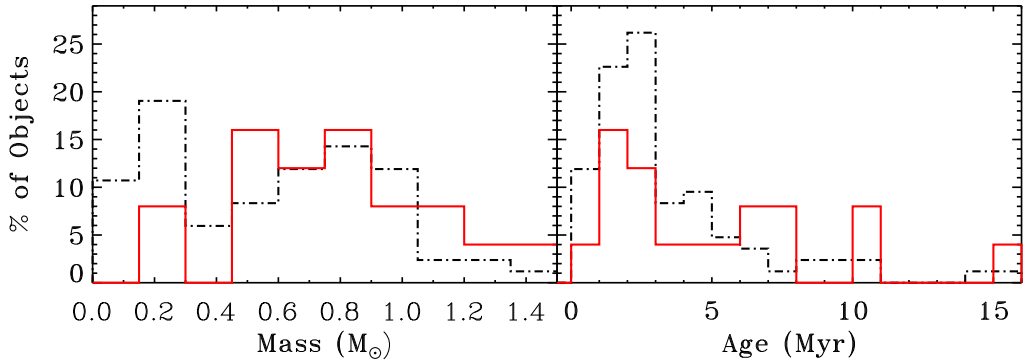} 
\end{center}
\caption{Distribution of masses and ages of the new young objects
  confirmed in Serpens (solid red lines). The distribution of the previously known
  YOSs in Serpens \citep{OL13} is shown for comparison (dot-dashed
  black lines).}
\label{hist}
\end{figure*}

Figure \ref{hist} shows the mass and age distributions for our sample
of newly-confirmed members of Serpens (solid red lines), and those for
the disk-bearing young stars in the same region (dot-dashed black
lines; \citealt{OL13}). It is immediately clear that, although not
perfect copies of each other (KS statistical tests show that the
hypothesis that they are drawn from different distributions is not
supported, with probabilities P = $24\%$ and $10\%$ for the
distributions of ages and masses, respectively), the distributions of
both stellar ages and masses of diskless young stars have the same
spread, and even the same peak locations, as those of disk-bearing
stars in the same region. Once again it is important to note that only
the confirmed new members are taken into consideration here. If the 4
older stars (Section \ref{shrd}) are indeed members of the cloud, the
statistical significance would change for the age distribution
(P=$0.09$ and $0.4$ for the distributions of ages and masses,
respectively). However, it is true that the bulk of this new
population is not consistently shifted towards larger ages.

Although we acknowledge that the sample studied here does not
necessarily represent the entire population of diskless young stars in
Serpens, the similarity of the distributions suggest that ages and
stellar masses are not absolute drivers of disk evolution. Other
parameters must play an important role, such that stars of similar age
and mass end up having such different disks (or lack thereof). That is
not to say that either parameter is not important for disk evolution:
the big picture remains that disks evolve with time and,
statistically, older stars have more evolved (or no) disks compared to
young stars. However, a complete theory of protoplanetary disk
evolution must take into account that stars of similar age and in the
same environment do present an large variety in disk structures (e.g
\citealt{FU06,FA09,OL10,OL13,SI11}) or no disks at all.

Similar results were seen by \citet{FM99}. From a reliable X-ray
sample from the ROSAT satellite, stars were characterized
spectroscopically and placed in HR diagrams. They report the in the
Chamaeleon I and Taurus-Auriga star-forming regions, no significant
differences are seen in the age distributions of disk-bearing (class
II) and diskless (class III) stars. In that work, the authors discuss
two immediate consequences of these results: (i) an underestimate of
cloud star formation rates for samples dominated by class II stars;
and (ii) that the high numbers of class III stars in young regions
imply that many (at least 1/2 of low-mass stars in Chamaeleon I) lose
their disks within 1 Myr. In that work, the authors already argued
that there is no preferred disk lifetime, since disk-bearing and
diskless stars coexist along the Hayashi track.

More recently, aided by {\it Chandra} observations and optical
spectroscopy, \citet{WI09,WI10} showed very similar age distributions
for class II and class III stars in the Serpens Core and NGC1333. They
argued that young stellar populations with and without disks show
indistinguishable spatial and age distributions, suggesting that class
III stars are not typically older stars, but rather stars that lost
their disks quicker.

In addition, \citet{CI07} and \citet{BE10} have studied the
differences between classical T Tauri stars (CTTS, young stars
actively accreting material from their disks) and weak-line T Tauri
stars (WTTS, young stars that show no signs of
accretion). Correspondingly, both studies find no significant
difference in the age distributions of CTTS and WTTS. However, it is
important to note that not all WTTS are bona-fide class III
objects. Many WTTS have disks (even thick disks) but that for some
reason are no longer accreting material onto their host
stars. Therefore, what these results show is that there is no
significant difference in the age distribution of stars that are
actively accreting from their disks and stars that are not.

Several different processes actively being studied could possibly be
responsible for the fast dissipation of some disks, while others last
many Myr. For instance, a relationship between the disk mass (from 1.3
mm observations with the Submillimeter Array) and stellar mass has
been shown to be real for a complete disk sample in Taurus by
\citet{AN13}. If confirmed for other complete and unbiased samples,
this result puts strong constraints on theories for disk evolution and
planet formation. The stellar mass would be decisive on how much
material is available in disks for planets to form from.

Another interesting process that could be decisive on disk evolution
timescales is photoevaporation.  Theoretical calculations predict that
energetic UV and X-ray photons emitted by the central star could heat
the surface of the disk and cause pressure-driven hydrodynamic mass
outflows from the disk
\citep{HL00,CL01,RA06a,RA06b,BE09,GO09,OW10}. Coupled with the disk
viscous evolution, disks could be dispersed in $\sim 10^6$
Myr. Observational confirmation of photoevaporation is an active field
that has already produced some interesting results
(e.g. \citealt{IP09,IP12}).

Just as exciting, and perhaps even more observationally challenging,
is the effect of multiplicity on disk evolution. Since the majority of
low-mass stars form in multiple systems, binary companions can affect
disk evolution directly. \citet{AK12} showed these effects for the
first time on a complete sample in Taurus-Auriga. They found that
tidal influence of a close binary ($\lesssim$ 40 AU) accelerates
significantly disk dispersal, with disks being completely dissipated
within 1 Myr for $\sim$2/3 of close binaries. The same is true for
only a small fraction of wide binaries and single stars ($\sim 10 -
20\%$) whose disks are dispersed within 2--3 Myr. These exciting
results must be confirmed with similar observations of complete
samples of young stars in different star-forming regions, of different
mean ages and environments.
\\

\section{Conclusions}
\label{scon}

We have confirmed, for the first time, a significant population of
young diskless stars discovered at X-ray wavelengths that are members
of the Serpens Molecular Cloud.

We have determined spectral types, effective temperatures and
luminosities using optical spectroscopy from VLT/FLAMES and
optical/near-IR photometry.

Stellar ages and masses for 23 stars were derived based on a
comparison with PMS evolutionary tracks. 19 young stars are confirmed
to be members of Serpens, while the remaining four could not be
conclusively shown to have ages consistent with cluster membership.
Six of these 19 confirmed stars show weak IR excess (below the Taurus
median), while the other 13 are truly diskless sources.

We find that the age and mass distributions of this new diskless
stellar population are remarkably similar to those of the disk-bearing
young stars in the same region, studied previously by \citet{OL13}.

This similarity may hold significant implications for the picture of
disk evolution in which the main factors that determine disk fraction
are mass and age. If confirmed for the larger sample of young diskless
star candidates, our results indicate that other, yet poorly
understood observationally, characteristics may play an important role
as well.

\acknowledgements This work is based on observations made with ESO
Telescopes at Paranal Observatories under program ID
083.C-0766(B). The authors would like to thank the anonymous referee
for her/his suggestions, that greatly improved this manuscript.

\clearpage

\begin{deluxetable}{c c c c}
\tabletypesize{\footnotesize}
\tablecolumns{4}
\tablewidth{0pt}
\tablecaption{Observational log \label{t_obs}} 
\tablehead{\colhead{Obs. Block ID} & 
           \colhead{Field}     &  
          \colhead{Obs. night}       &  
           \colhead{Exp. time (s)} 
}
\startdata
357791             & Serpens AO5 &  11/Jun/2009   & 2000          \\
357791             & Serpens AO5 &  11/Jun/2009   & 2000          \\
357792             & Serpens AO5 &  11/Jun/2009   & 2000          \\
357792             & Serpens AO5 &  11/Jun/2009   & 2000          \\
357792             & Serpens AO5 &  11/Jun/2009   & 2000          \\
357793             & Serpens AO6 &  08/Jun/2009   & 2000          \\
357793             & Serpens AO6 &  08/Jun/2009   & 2000          \\
365063             & Serpens AO6 &  11/Jun/2009   & 2000          \\
365063             & Serpens AO6 &  11/Jun/2009   & 2000          \\
\enddata
\end{deluxetable}

\begin{deluxetable}{l l l l l l l l l c}
\tabletypesize{\tiny}
\tablecolumns{10}
\tablewidth{0pt}
\tablecaption{Observed objects in field AO5 \label{t_radec5}}
\tablehead{\colhead{ID\tablenotemark{a}} &
             \colhead{RA (deg)}         & 
             \colhead{DEC (deg)}       &
             \colhead{2MASS Name} &
             \colhead{R (mag)}           &
             \colhead{Z (mag)}           &
             \colhead{J (mag)}            &
             \colhead{H (mag)}          & 
             \colhead{K (mag)}           &
           \colhead{Good S/N?}
}
\startdata
03-005 &     277.39835 &       0.584499 & 18293563+0035035 & 18.2 & 14.3 & 12.0 & 10.8 & 10.3 & Yes \\
03-007 &     277.29083 &       0.492694 & 18290985+0029329 & 19.2 & 15.6 & 12.4 & 10.8 & 10.1 & No \\
03-008 &     277.24176 &       0.290611 & 18285808+0017243 & 21.9 & 18.0 & 10.7 &  9.8 &  9.4 & No \\
03-009 &     277.25363 &       0.492055 & 18290088+0029312 & 17.0 & 14.1 & 11.6 & 10.0 &  9.0 & Yes \\
03-011 &     277.37317 &       0.558027 & 18292959+0033280 & 20.8 & 16.0 & 13.7 & 12.6 & 12.1 & No \\
03-012 &     277.37091 &       0.302722 & 18292905+0018091 & 17.2 & 14.7 & 12.4 & 11.4 & 10.9 & Yes \\
03-013 &     277.28870 &       0.304611 & 18290934+0018159 & 22.4 & 18.3 & 14.4 & 12.0 & 11.0 & No \\
03-014 &     277.27942 &       0.643972 & 18290698+0038378 & 21.9 & 18.0 & 15.2 & 12.9 & 11.5 & No \\
03-015 &     277.43549 &       0.565333 & 18294460+0033553 & 19.4 & 15.7 & 12.7 & 11.9 & 11.5 & No \\
03-017 &     277.34201 &       0.672361 & 18292209+0040203 & 23.0 & 17.4 & 13.3 & 11.1 & 10.0 & Yes \\
03-020 &     277.26788 &       0.556611 & 18290437+0033235 & 21.9 & 18.0 & 15.9 & 13.0 & 11.3 & No \\
03-021 &     277.35974 &       0.501416 & 18292641+0030043 & 21.9 & 18.0 & 16.2 & 13.9 & 12.9 & No \\
03-022 &     277.25016 &       0.282972 & 18290025+0016578 & 17.7 & 15.2 & 12.9 & 11.5 & 10.7 & No \\
03-023 &     277.32410 &       0.297277 & 18291770+0017488 & 21.9 & 18.0 & 13.3 & 12.7 & 12.4 & No \\
03-025 &     277.40137 &       0.565749 & 18293634+0033559 & 21.4 & 16.3 & 13.8 & 12.3 & 11.6 & No \\
03-026 &     277.28122 &       0.456777 & 18290748+0027234 & 17.0 & 14.7 & 13.2 & 12.6 & 12.3 & Yes \\
03-027 &     277.31396 &       0.658166 & 18291521+0039343 & 20.4 & 16.1 & 13.4 & 12.1 & 11.5 & Yes \\
03-028 &     277.17624 &       0.627361 & 18283136+0030534 & 21.9 & 18.0 & 14.4 & 13.0 & 12.3 & No \\
03-029 &     277.27597 &       0.329388 & 18290615+0019442 & 18.8 & 15.6 & 13.4 & 12.2 & 11.8 & No \\
03-030 &     277.43771 &       0.540333 & 18294516+0032245 & 20.3 & 16.6 & 13.2 & 11.9 & 11.3 & No \\
03-032 &     277.24292 &       0.436444 & 18285828+0026103 & 17.8 & 15.5 & 14.0 & 13.4 & 13.2 & Yes \\
03-033 &     277.31750 &       0.306361 & 18291617+0018222 & 21.0 & 18.0 & 14.3 & 11.8 & 10.1 & No \\
03-034 &     277.13470 &       0.493916 & -- & 21.9 & 18.0 &  -- &  -- &  -- & No \\
03-035 &     277.26584 &       0.339027 & 18290394+0020212 & 15.6 & 13.7 & 11.6 & 10.1 &  9.1 & No \\
03-037 &     277.43726 &       0.590944 & 18294502+0035264 & 21.6 & 18.3 & 14.1 & 12.3 & 11.4 & No \\
03-038 &     277.12064 &       0.588888 & 18282871+0035206 & 19.8 & 16.8 & 14.2 & 13.0 & 12.5 & No \\
\enddata
\tablenotetext{a}{As in Brown et al., in prep.}
\end{deluxetable}

\begin{deluxetable}{l l l l l l l l l c}
\tabletypesize{\tiny}
\tablecolumns{10}
\tablewidth{0pt}
\tablecaption{Observed objects in field AO6 \label{t_radec6}} 
\tablehead{\colhead{ID\tablenotemark{a}} & 
             \colhead{RA (deg)}         & 
             \colhead{DEC (deg)}       &
             \colhead{2MASS Name} &
             \colhead{R (mag)}           &
             \colhead{Z (mag)}           &
             \colhead{J (mag)}            &
             \colhead{H (mag)}          & 
             \colhead{K (mag)}           &
           \colhead{Good S/N?}
}
\startdata
02-002 &     277.50937 &       0.940555 & 18300227+0056259 & 13.5 & 12.3 & 10.8 & 10.1 &  9.9 & Yes \\
02-004 &     277.34213 &       0.672472 & 18292209+0040203 & 23.0 & 17.4 & 13.3 & 11.1 & 10.0 & No \\
02-005 &     277.43216 &       0.808805 & 18294372+0048308 & 15.8 & 14.7 & 13.3 & 12.6 & 12.5 & Yes \\
02-006 &     277.22983 &       0.913722 & 18285512+0054496 & 19.7 & 16.7 & 14.2 & 13.0 & 12.6 & No \\
02-007 &     277.18753 &       0.756916 & 18284499+0045239 & 16.6 & 14.0 & 11.9 & 10.8 & 10.4 & Yes \\
02-008 &     277.51935 &       0.707138 & 18300459+0042247 & 16.2 & 14.3 & 12.5 & 11.5 & 11.2 & Yes \\
02-009 &     277.31503 &       0.656249 & 18291564+0039227 & 21.1 & 17.0 & 14.1 & 12.8 & 12.1 & No \\
02-010 &     277.54959 &       0.781416 & 18301185+0046519 & 17.0 & 15.2 & 13.4 & 12.5 & 12.1 & No \\
02-013 &     277.47601 &       0.752222 & 18295424+0045074 & 13.5 & 11.9 & 10.4 &  9.8 &  9.5 & Yes \\
02-014 &     277.52570 &       0.709277 & 18300616+0042336 & 14.2 & 12.3 &  9.8 &  8.6 &  7.5 & Yes \\
02-017 &     277.21396 &       0.873749 & 18285134+0052270 & 16.1 & 13.5 & 11.9 & 11.3 & 11.0 & No \\
02-019 &     277.45401 &       0.731916 & -- & -- & -- & -- &  -- &  -- & No \\
02-021 &     277.22470 &       0.765194 & 18285396+0045528 & 18.7 & 15.7 & 13.3 & 12.1 & 11.5 & No \\
02-022 &     277.35745 &       0.722611 & -- & -- & -- &  -- &  -- &  -- & No \\
02-024 &     277.35950 &       0.875499 & -- & -- & -- &  -- &  -- &  -- & No \\
02-025 &     277.56018 &       0.876055 & 18301440+0052352 & 17.0 & 15.2 & 12.5 & 11.6 & 11.2 & Yes \\
02-027 &     277.50992 &       0.869249 & 18300236+0052098 & 17.6 & 15.1 & 13.1 & 12.0 & 11.6 & Yes \\
02-028 &     277.45834 &       0.850444 & 18295003+0051014 & 18.0 & 15.3 & 13.0 & 11.9 & 11.4 & Yes \\
02-029 &     277.42606 &       0.828666 & 18294221+0049431 & 19.5 & 16.5 & 14.0 & 12.8 & 12.3 & Yes \\
02-030 &     277.48447 &       0.657055 & 18295628+0039246 & 17.0 & 15.2 & 13.1 & 12.1 & 11.7 & No \\
02-031 &     277.42199 &       0.817277 & 18294122+0049020 & 19.3 & 16.2 & 13.8 & 12.5 & 11.9 & No \\
02-033 &     277.28818 &       0.823861 & -- & -- & -- &  -- &  -- &  -- & No \\
02-036 &     277.25601 &       0.851166 & -- & 21.2 & 18.8 &  -- &  -- &  -- & No \\
02-037 &     277.44699 &       0.665388 & 18294727+0039555 & 17.0 & 15.2 & 13.6 & 12.6 & 12.1 & Yes \\
02-038 &     277.46878 &       0.649944 & -- & -- & -- &  -- &  -- &  -- & No \\
02-039 &     277.41428 &       0.715833 & 18293946+0042570 & 18.6 & 16.6 & 13.6 & 12.4 & 11.9 & Yes \\
02-041 &     277.43964 &       0.934277 & 18294546+0056026 & 19.6 & 16.4 & 13.5 & 12.1 & 11.4 & No \\
02-043 &     277.26514 &       0.753805 & 18290362+0045135 & 19.9 & 16.4 & 13.5 & 12.1 & 11.6 & Yes \\
02-047 &     277.35171 &       0.805138 & -- & -- & -- &  -- &  -- &  -- & No \\
02-048 &     277.30933 &       0.787027 & 18291396+0047165 & 20.8 & 17.8 & 15.1 & 13.8 & 13.4 & No \\
02-049 &     277.40204 &       0.775999 & 18293650+0046333 & 19.5 & 16.2 & 13.5 & 12.2 & 11.7 & Yes \\
02-050 &     277.28921 &       0.775749 & 18290943+0046343 & 18.1 & 15.3 & 13.3 & 12.2 & 11.8 & No \\
02-051 &     277.55670 &       0.784777 & 18301365+0047055 & 17.0 & 15.2 & 13.2 & 12.3 & 12.0 & Yes \\
02-052 &     277.48480 &       0.835527 & -- & -- & -- &  -- &  -- &  -- & No \\
02-053 &     277.51263 &       0.879194 & 18300298+0052460 & 19.8 & 16.8 & 14.5 & 13.7 & 13.2 & No \\
02-054 &     277.35162 &       0.938861 & 18292426+0056189 & 17.6 & 14.8 & 12.5 & 11.5 & 11.1 & No \\
02-055 &     277.34671 &       0.859694 & -- & -- & -- &  -- &  -- &  -- & No \\
02-056 &     277.24307 &       0.934805 & 18285837+0056050 & 20.4 & 16.6 & 14.1 & 13.1 & 12.6 & No \\
02-057 &     277.40082 &       0.704499 & 18293620+0042163 & 18.8 & 16.6 & 13.3 & 11.9 & 11.3 & Yes \\
02-058 &     277.30011 &       0.820249 & -- & 22.7 & 20.4 &  -- &  -- &  -- & No \\
02-059 &     277.26154 &       0.767055 & -- & -- & -- &  -- &  -- &  -- & No \\
02-060 &     277.20172 &       0.798305 & 18284840+0047515 & 17.2 & 15.1 & 13.2 & 12.2 & 11.9 & No \\
02-061 &     277.53561 &       0.811666 & -- & 22.2 & 19.2 &  -- &  -- &  -- & No \\
02-062 &     277.35211 &       0.922777 & -- & -- & -- &  -- &  -- &  -- & No \\
02-063 &     277.38480 &       0.900916 & 18293237+0054024 & 18.0 & 15.6 & 13.5 & 12.5 & 12.1 & Yes \\
02-064 &     277.45538 &       0.775388 & -- & -- & -- &  -- &  -- &  -- & No \\
02-065 &     277.47226 &       0.914527 & 18295337+0054523 & 19.7 & 16.7 & 14.0 & 12.7 & 12.1 & No \\
02-068 &     277.34466 &       0.815555 & 18292277+0048569 & 19.5 & 16.6 & 14.0 & 12.8 & 12.4 & No \\
02-069 &     277.40829 &       0.888111 & 18293799+0053154 & 22.7 & 18.9 & 15.6 & 14.3 & 13.7 & No \\
02-072 &     277.53726 &       0.789472 & 18300893+0047219 & 17.7 & 15.4 & 13.5 & 12.4 & 12.0 & Yes \\
02-073 &     277.27920 &       0.643722 & 18290698+0038378 & 17.0 & 15.2 & 15.2 & 12.9 & 11.5 & No \\
02-074 &     277.42184 &       0.962361 & -- & 22.3 & 19.6 &  -- &  -- &  -- & No \\
02-076 &     277.35800 &       0.744583 & 18292588+0044396 & 20.4 & 16.2 & 13.7 & 12.5 & 12.2 & No \\
02-079 &     277.39325 &       0.951666 & -- & -- & -- &  -- &  -- &  -- & No \\
02-080 &     277.44568 &       0.726527 & 18294682+0043396 & 17.0 & 15.2 & 15.7 & 13.6 & 12.7 & No \\
02-081 &     277.39862 &       0.714861 & 18293575+0042499 & 22.8 & 19.0 & 15.5 & 14.2 & 13.5 & No \\
02-083 &     277.46017 &       0.728611 & 18295042+0043435 & 17.8 & 15.4 & 13.0 & 11.5 & 10.5 & No \\
02-086 &     277.40628 &       0.748555 & -- & -- & -- &  -- &  -- &  -- & No \\
02-088 &     277.38766 &       0.669555 & 18293301+0040087 & 17.0 & 15.2 & 14.5 & 12.9 & 12.0 & No \\
02-093 &     277.44122 &       0.717305 & -- & -- & -- &  -- &  -- &  -- & No \\
\enddata
\tablenotetext{a}{As in Brown et al., in prep.}
\end{deluxetable}

\begin{deluxetable}{l c c c l c c c}
\tabletypesize{\footnotesize}
\tablecolumns{8}
\tablewidth{0pt}
\tablecaption{Spectral types of objects with classifiable spectra \label{t_spt}} 
\tablehead{\colhead{Object ID} & 
           \colhead{SpT}     &  
           \colhead{SpT range}       &  
           \colhead{H$\alpha$\tablenotemark{a}}     &  
           \colhead{Li \tablenotemark{a}}       &  
           \colhead{IR excess?}     &
           \colhead{X-ray Counts/s\tablenotemark{b,c}}       &  
           \colhead{Log L$_X$ (erg/s)\tablenotemark{b}}      
}
\startdata
03-005   & M0    &  M0.5 -- K0   & strong E & A & Yes &    0.0367(0.0303-0.0444)     &   30.05\\
03-009   & M0.5  &  M2 -- K7     & strong E & A &        &   0.0193(0.0151-0.0243)      &  29.62\\
03-012   & M0    &  M2 -- K2      & weak E & A &       &    0.0321(0.0244-0.0416)      &  29.95\\    
03-017   & K5     &  K7 -- G7      & A & -- &     &    0.0275(0.0207-0.0361)      &  29.79\\
03-026   & M4    &  M6 -- M2     & weak E & -- &      &    0.0048(0.0018-0.0086)      &  29.09 \\
03-027   & G8     &  K5 -- G0      & A & -- &     &    0.0075(0.0049-0.0109)      &  29.47\\
03-032   & M4     &  M6 -- M2    & weak E & -- &      &    0.0038(0.0007-0.0079)      &  28.86 \\
02-002   & K2     &  K7 -- K0      & A & A &       &    0.0908(0.0845-0.0976)      &  30.35 \\
02-005   & M0    &  M0.5 -- K5  & weak E & -- &       &    0.0197(0.0175-0.0223)      &  29.70\\
02-007   & M2    &  M3 -- M0.5 & weak E & A & Yes   &    0.0569(0.0508-0.0639)     &   30.10\\
02-008   & K2     &  K7 -- K0     & -- & A &        &    0.0338(0.0299-0.0382)     &   29.90\\
02-013   & G5     &  G8 -- G2.5 & A & A &          &    0.0150(0.0127-0.0177)     &   29.52\\
02-014   & G7     &  K0 -- F0     & weak E & A &          &    0.0184(0.0154-0.0219)     &   29.60  \\         
02-025   & G5     &  G8 -- G2.5  & A & -- &       &    0.0103(0.0084-0.0126)     &   29.84\\
02-027   & M2    &  M3 -- M0.5 & weak E & A &          &    0.0112(0.0091-0.0139)     &   29.41\\
02-028   & M3    &  M4 -- M0.5 & weak E & A & Yes    &    0.0073(0.0058-0.0092)    &    29.29\\
02-029   & M0    &  M0.5 -- K7  & weak E & A &          &    0.0065(0.0051-0.0081)     &   29.20\\
02-037   & M2    &  M3 -- M0     & strong E & A & Yes   &    0.0073(0.0053-0.0098)     &   29.27\\
02-039   & K5     &  K7 -- K0      & -- & A &        &    0.0079(0.0062-0.0100)     &   29.32\\
02-043   & M0.5 &  M3 -- K7     & weak E & A &         &    0.0079(0.0060-0.0102)     &   29.24  \\  
02-049   & M2    &  M3 -- M0.5 & weak E & A &           &    0.0042(0.0030-0.0056)     &   28.92\\
02-051   & M0.5 &  M2 -- K7     & weak E & A &         &    0.0069(0.0049-0.0094)     &   29.27\\
02-057   & K5     &  K7 -- K0      & A & A & Yes   &    0.0062(0.0046-0.0082)      &  29.13 \\
02-063   & M0.5 &  M3 -- M0    & weak E & A &           &    0.0032(0.0020-0.0048)     &   28.92\\
02-072   & M2    &  M3 -- M0.5 & strong E & A & Yes     &    0.0051(0.0034-0.0073)     &   29.11\\
\enddata
\tablenotetext{a}{E denotes emission and A denotes absorption}
\tablenotetext{b}{From Brown et al., in prep.}
\tablenotetext{c}{In parenthesis is the 95\% confidence range}
\end{deluxetable}

\begin{deluxetable}{l l l l l l}
\tabletypesize{\footnotesize}
\tablecolumns{6}
\tablewidth{0pt}
\tablecaption{Stellar Parameters for the young diskless stars
  in Serpens \label{t_res}} 
\tablehead{\colhead{Object ID} & 
           \colhead{Av (mag)}     &  
           \colhead{T$_{eff}$ (K)}       &  
           \colhead{L$_{*}$ (L$_{\odot}$)}     &  
           \colhead{Mass (M$_{\odot}$)}       &  
           \colhead{Age (Myr)} 
}
\startdata
03-005  & 4.6    &  3850$^{+1400}_{-65}$  &   1.52$^{+1.22}_{-0.68}$ &     1.06$^{+0.18}_{-0.08}$ &     1.01$^{+20.31}_{-1.01}$  \\ 
03-012  &  3.2   &  3850$^{+1050}_{-290}$ &   0.73$^{+0.58}_{-0.33}$ &     0.98$^{+0.06}_{-0.33}$ &     3.06$^{+27.13}_{-1.94}$  \\ 
03-017  &  13.6 &  4350$^{+1280}_{-290}$ &   5.87$^{+4.72}_{-2.64}$ &     1.60$^{+0.12}_{-1.60}$ &     1.00$^{+6.74}_{-1.00}$    \\ 
03-026  &  0.0   &  3270$^{+290}_{-280}$  &   0.13$^{+0.11}_{-0.06}$ &     0.26$^{+0.25}_{-0.15}$ &     2.67$^{+7.06}_{-2.67}$  \\ 
03-032  &  0.0   &  3270$^{+290}_{-280}$  &   0.06$^{+0.05}_{-0.03}$ &     0.23$^{+0.27}_{-0.14}$ &     6.83$^{+23.38}_{-5.73}$  \\ 
02-002  &  2.0   &  4900$^{+350}_{-840}$  &   3.01$^{+2.42}_{-1.35}$ &     1.79$^{+0.31}_{-1.10}$ &     3.95$^{+3.79}_{-3.95}$  \\ 
02-005  &  0.0   &  3850$^{+500}_{-65}$   &   0.13$^{+0.11}_{-0.06}$ &     0.73$^{+0.09}_{-0.13}$ &   30.71$^{+47.73}_{-27.23}$ \\ 
02-007  &  3.4   &  3560$^{+225}_{-145}$  &   1.14$^{+0.91}_{-0.51}$ &     0.75$^{+0.22}_{-0.13}$ &     0.41$^{+1.03}_{-0.41}$ \\ 
02-008  &  4.0   &  4900$^{+350}_{-840}$  &   1.09$^{+0.87}_{-0.49}$ &     1.18$^{+0.00}_{-1.18}$ &   20.16$^{+22.94}_{-17.29}$\\ 
02-013  &  3.2   &  5770$^{+75}_{-250}$    &   8.22$^{+6.60}_{-3.69}$ &     1.84$^{+0.40}_{-0.35}$ &     6.69$^{+5.10}_{-2.55}$ \\ 
02-014  &  10.3 &  5630$^{+1570}_{-380}$ & 86.46$^{+69.45}_{-38.86}$ &    3.35$^{+0.09}_{-0.09}$ &   1.71$^{+0.44}_{-0.44}$ \\ 
02-025  &  5.1   &  5770$^{+75}_{-250}$    &   1.92$^{+1.54}_{-0.86}$ &     1.35$^{+0.01}_{-1.35}$ &   24.50$^{+980.15}_{-15.24}$ \\ 
02-027  &  2.8   &  3560$^{+225}_{-145}$   &   0.34$^{+0.27}_{-0.15}$ &     0.55$^{+0.26}_{-0.13}$ &     2.93$^{+4.51}_{-1.42}$  \\ 
02-028  &  3.5   &  3415$^{+370}_{-145}$   &   0.40$^{+0.32}_{-0.18}$ &     0.45$^{+0.38}_{-0.14}$ &     1.44$^{+4.56}_{-0.99}$  \\ 
02-029  &  4.6   &  3850$^{+210}_{-65}$    &   0.23$^{+0.19}_{-0.11}$ &     0.81$^{+0.09}_{-0.09}$ &    15.32$^{+16.65}_{-8.32}$  \\ 
02-037  &  2.8   &  3560$^{+290}_{-145}$   &   0.21$^{+0.17}_{-0.10}$ &     0.53$^{+0.27}_{-0.14}$ &     5.68$^{+11.80}_{-3.10}$  \\ 
02-039  &  5.1   &  4350$^{+900}_{-290}$   &   0.49$^{+0.39}_{-0.22}$ &     1.03$^{+0.25}_{-0.18}$ &   19.55$^{+638.39}_{-11.23}$ \\ 
02-043  &  5.4   &  3785$^{+275}_{-370}$   &   0.47$^{+0.38}_{-0.21}$ &     0.85$^{+0.20}_{-0.37}$ &     4.71$^{+4.98}_{-3.58}$  \\ 
02-049  &  4.8   &  3560$^{+225}_{-145}$   &   0.39$^{+0.31}_{-0.17}$ &     0.57$^{+0.26}_{-0.12}$ &     2.59$^{+3.67}_{-1.38}$  \\ 
02-051  &  1.9   &  3785$^{+275}_{-225}$   &   0.25$^{+0.20}_{-0.11}$ &     0.78$^{+0.07}_{-0.25}$ &    10.36$^{+15.54}_{-5.69}$  \\ 
02-057  &  6.6   &  4350$^{+900}_{-290}$   &   0.96$^{+0.77}_{-0.43}$ &     1.31$^{+\dagger}_{-1.31}$ &     7.56$^{+35.04}_{-4.61}$  \\ 
02-063  &  3.0   &  3785$^{+ 65}_{-370}$    &   0.24$^{+0.19}_{-0.11}$ &     0.78$^{+0.06}_{-0.39}$ &    10.67$^{+15.66}_{-8.16}$ \\ 
02-072  &  3.0   &  3650$^{+135}_{-235}$   &   0.24$^{+0.19}_{-0.11}$ &     0.63$^{+0.15}_{-0.23}$ &     7.21$^{+9.11}_{-4.65}$  \\
\enddata
\end{deluxetable}

\label{lastpage}

\end{document}